\documentclass[twocolumn,prb,amsmath,amssymb,amsfonts,superscriptaddress,floatfix,aps,showpacs,notitlepage]{revtex4-1}

\usepackage{graphicx}
\usepackage{epstopdf}
\usepackage{dcolumn}
\usepackage{bm}
\usepackage{rotating} 
\usepackage{color}
\usepackage{subfigure}
\usepackage{natbib}
\usepackage{enumitem}
\usepackage[Symbol]{upgreek}
\usepackage[symbol*]{footmisc}
\usepackage{lipsum}
\usepackage{amsmath}                    
\usepackage{amssymb}                    

\usepackage{textcomp}		
\usepackage{tabularx}
\usepackage{float}		
\usepackage{multirow}	

\begin{document}
\newcommand{\oscar}[1]{\textcolor{red} {#1}}
\renewcommand{\thetable}{\arabic{table}}

\title{An ab-initio study of the electron-phonon coupling within a Cr(001)-surface}

\date{\today}

\author{L. Peters}
\email{L.Peters@science.ru.nl}
\affiliation{Institute for Molecules and Materials, Radboud University 
Nijmegen, NL-6525 AJ Nijmegen, The Netherlands}

\author{A. N. Rudenko}
\affiliation{School of Physics and Technology, Wuhan University, Wuhan 430072, China}
\affiliation{Theoretical Physics and Applied Mathematics Department, Ural Federal
University, 620002 Ekaterinburg, Russia}
\affiliation{Institute for Molecules and Materials, Radboud University 
Nijmegen, NL-6525 AJ Nijmegen, The Netherlands}

\author{M. I. Katsnelson}
\affiliation{Institute for Molecules and Materials, Radboud University Nijmegen, NL-6525 AJ Nijmegen, 
The Netherlands}
\affiliation{Theoretical Physics and Applied Mathematics Department, Ural Federal
University, 620002 Ekaterinburg, Russia}

\begin{abstract}
It is experimentally well established that the Cr(001)-surface exhibits a sharp resonance around the Fermi level. However, there is no consensus about its physical origin. It is proposed to be either due to a single particle \textit{$d_{z^{2}}$} surface state renormalised by electron-phonon coupling or the orbital Kondo effect involving the degenerate \textit{$d_{xz}$}/\textit{$d_{yz}$} states. In this work we examine the electron-phonon coupling of the Cr(001)-surface by means of \textit{ab-initio} calculations in the form of density functional perturbation theory. More precisely, the electron-phonon mass-enhancement factor of the surface layer is investigated for the 3d states. For the majority and minority spin \textit{$d_{z^{2}}$} surface states we find values of $0.19$ and $0.16$. We show that these calculated electron-phonon mass-enhancement factors are not in agreement with the experimental data even if we use realistic values for the temperature range and surface Debye frequency for the fit of the experimental data. More precisely, then experimentally an electron-phonon mass-enhancement factor of $0.70\pm0.10$ is obtained, which is not in agreement with our calculated values of $0.19$ and $0.16$. Therefore, we conclude that the experimentally observed resonance at the Cr(001)-surface is not due to polaronic effects, but due to electron-electron correlation effects.  
\end{abstract}

\pacs{}
\maketitle


\section{Introduction}
The electrode-electrolyte interface in a battery, topological insulators, multi-layer devices and at contacts are examples where surface physics plays an important role. In the field of spintronics magnetic multi-layers are used to exploit the different tunneling probabilities of the spin-up and spin-down electrons for the development of novel devices. For this purpose a thorough understanding of the physical mechanisms behind the tunneling process is required.~\cite{tun1,tun2} For example, in chromium magnetic multi-layers it is known that complicated many-body effects at the surface are responsible for the tunneling.~\cite{tun3}

Surface physics is also fundamentally very interesting due to the occurence of new and unexpected features. For example, for the Cr(001)-surface a sharp resonance close to the Fermi level is observed in angular resolved photoemission and scanning tunneling experiments.~\cite{arpes1,arpes2,arpes3,dzsur1} Another reason for the large interest in chromium is its peculiar magnetic properties. Its magnetic ground state is described by a spin-density wave with a long period modulating the amplitude of the magnetic moments along the $\langle001\rangle$ direction, which is incommensurate with the underlying body centered cubic structure.~\cite{sdw}

In order to understand the physical origin of the observed resonance at the Fermi level several experimental and theoretical investigations were conducted. The first idea was to explain the resonance in terms of a single particle \textit{$d_{z^{2}}$} surface state.~\cite{dzsur1,okon6} A shortcoming of this idea was the unrealistic reduction of the magnetic moment required to obtain the correct resonance position. Another interpretation in terms of an orbital Kondo effect involving the degenerate \textit{$d_{xz}$} and \textit{$d_{yz}$} states was proposed to explain the scanning tunneling spectroscopy experiments on very clean Cr(001)-surfaces.~\cite{okon1,okon2} Then, scanning tunneling spectroscopy experiments were performed in a wide temperature range.~\cite{okon3} It was observed that the experimental data could be explained both in terms of the \textit{$d_{z^{2}}$} surface state and orbital Kondo effect. Although for the former an electron-phonon coupling strength 5-10 times larger than in the bulk had to be assumed. On the other hand the orbital Kondo effect was called into question by a combination of scanning tunneling spectroscopy, photoemission spectroscopy and inverse photoemission spectroscopy experiments.~\cite{okon4} It was demonstrated that the resonance above the Fermi level was mainly of \textit{$d_{z^{2}}$} character. However, it should be realized that the resolution of inverse photoemission spectroscopy is not sufficient to be conclusive about the character of the sharp resonance emerging at low temperatures.

In the newest experiments a different behavior is observed than in the earlier experiments.~\cite{okon5} Namely a pseudogap is found below roughly 200~K and the emergence of a sharp resonance below 75~K. These observations hint in the direction of a many-body interpretation of the resonance just as recent dynamical mean-field theory (DMFT) calculations do.~\cite{schuler,larscr} For example, based on DMFT calculations within the continuous-time quantum Monte Carlo (CTQMC) solver it was observed that the resonance was very robust against artificial shifts in the one-particle energies of the \textit{$d_{xz}$}, \textit{$d_{yz}$} and \textit{$d_{z^{2}}$} states. In contrast for DMFT calculations based on the spin-polarized T-matrix fluctuation exchange approximation, which is suitable for weakly and moderately correlated systems, no resonance was observed. However, the high-energy features, everything except the resonance, of the spectrum were successfully explained within this method. Finally, the non-crossing approximation (NCA) was employed within DMFT, which is basically designed to capture (orbital) Kondo like processes. Depending on the starting point, it was observed that an orbital Kondo resonance might evolve in the presence of a strong magnetic field like in the Cr(001)-surface. This could not be verified due to spurious behavior involved with the spin-polarized version of the NCA method.~\cite{larscr,kroh}

In this work it is investigated whether the $d_{z^2}$ surface state renormalized by means of the electron-phonon coupling is responsible for the experimentally observed resonance at the Fermi level of Cr(001)-surfaces. More precisely, the electron-phonon mass-enhancement factor of the Cr(001)-surface is investigated and a comparison is made with that of the bulk by means of ab-initio calculations. For this purpose a linear response scheme in terms of the density functional perturbation theory (DFPT) within a pseudopotential plane-wave approach is employed.~\cite{dfpt1,dfpt2,malgor} We first tested this method on paramagnetic and anti-ferromagnetic Cr-bulk and found electron-phonon mass-enhancement factors in reasonable agreement with strong-coupling theory and good agreement with optical pump-and-probe experiment respectively.~\cite{optexp,strongt} For the Cr(001)-surface we obtained for the majority and minority spin \textit{$d_{z^{2}}$} surface states electron-phonon mass-enhancement factors of respectively $0.19$ and $0.16$. We show that these calculated values are not in agreement with the experimental data even if we use realistic values for the temperature range and surface Debye frequency for the fit of the experimental data. More precisely, then experimentally an electron-phonon mass-enhancement factor of $0.70\pm0.10$ is obtained, which is not in agreement with our calculated values of $0.19$ and $0.16$. Therefore, we conclude that the experimentally observed resonance at the Cr(001)-surface is not due to polaronic effects, but due to electron-electron correlation effects. More studies are needed to exactly determine which many-body processes are responsible for the resonance.

The rest of the paper is organized as follows. The method and computational details are presented in Section~\ref{section-2}. Section~\ref{section-3} contains the results and discussion and finally in Section~\ref{section-4} we conclude.

\section{Method and computational details} 
\label{section-2}
The response of an electron system to external perturbations is commonly studied in physics. An efficient and accurate technique to do this is density-functional perturbation theory (DFPT), which is a combination of density-functional theory (DFT) and linear response theory.~\cite{dfpt1,dfpt2,dft1,dft2} For example, this method allows the investigation of the coupling between the electrons and phonons in a system. It is known that many physical properties are determined by this coupling, e.g. electrical and thermal conductivity, and supercondutivity. 

DFT is based on the fact that the total energy of an interacting electron system is a functional of the electron density and the variational principle.~\cite{dft1} In order to obtain the ground state of the interacting electron system in general a mapping to a set of single-particle equations is performed,~\cite{dft2}
\begin{equation}
\begin{gathered}
\hat{H}_{DFT}\psi_{{\bf k}\nu} = \Bigg( \frac{{\bf \hat{p}}^2}{2m} + V_{eff}[n] \Bigg) \psi_{{\bf k}\nu} = \epsilon_{{\bf k}\nu} \psi_{{\bf k}\nu},\\
V_{eff}[n]=V_{ion}[n]+V_{H}[n]+V_{xc}[n]. 
\end{gathered}
\label{dft1}
\end{equation}
\newline
Here the first term between brackets is the kinetic energy operator. The $\psi_{{\bf k}\nu}$ and $\epsilon_{{\bf k}\nu}$ are the so called Kohn-Sham eigenstates and eigenenergies. Further, $V_{eff}$ is the effective potential which is a functional of the electron density. This functional can be separated into three parts, the interaction of the electrons with the ions ($V_{ion}$), a Hartree term ($V_{H}$) and exchange-correlation part ($V_{xc}$). The latter contains the exchange and correlation effects among the electrons. For this part the functional dependence on the density is not exactly known and approximations are used in practice. The most popular approximations are derived in the limit of a (nearly) uniform electron gas, i.e. the local density approximation (LDA) and the generalized gradient approximation (GGA).~\cite{lda1,lda2,gga} Then, by using the following expression for the electron density,
\begin{equation}
\begin{gathered}
n({\bf r})= \sum_{{\bf k}\nu} f_{{\bf k}\nu} |\psi_{{\bf k}\nu}|^{2}. 
\end{gathered}
\label{dft2}
\end{equation}
\newline
the system can be solved self-consistently. Here, $f_{{\bf k}\nu}$ corresponds to the occupation number of the state $\psi_{{\bf k}\nu}$. 

After obtaining the self-consistent solution of Eq.~\ref{dft1} and~\ref{dft2}, linear response theory can be used to investigate the coupling of the phonon and electron systems. For this purpose the displacement pattern corresponding to a phonon is considered as a static perturbation for the electron system within DFPT, i.e. the Born-Oppenheimer approximation. As can be inferred from Eq.~\ref{dft1} a perturbation will lead to a change of the electron density $\Delta n$ and effective potential $\Delta V_{eff}$, which within linear response can be obtained self-consistently from the following set of equations
\begin{equation}
\begin{gathered}
\Delta n= 4Re\sum_{{\bf k}\nu}^{occ} \psi_{{\bf k}\nu}^{\ast} \Delta\psi_{{\bf k}\nu}, \\
\big( \hat{H}_{DFT} -\epsilon_{{\bf k}\nu} \big) \Delta \psi_{{\bf k}\nu} = - \big(\Delta V_{eff} -\Delta \epsilon_{{\bf k}\nu} \big) \psi_{{\bf k}\nu}, \\
\Delta V_{eff} = \Delta V_{ion} + \frac{1}{2}\int \frac{\Delta n({\bf r'})}{|{\bf r} - {\bf r'}|}d{\bf r'} + \frac{dV_{xc}}{dn}\bigg|_{n}\Delta n.
\end{gathered}
\label{dft3}
\end{equation}
\newline
Here the tag \textit{occ} above the summation correponds to a summation over occupied states only and $\Delta \epsilon_{{\bf k}\nu}=\langle \psi_{{\bf k}\nu} | \Delta V_{eff}| \psi_{{\bf k}\nu} \rangle$ is the first-order variation of the Kohn-Sham eigenvalue. 

After the set of equations in Eq.~\ref{dft3} has been solved self-consistently for the atomic perturbations, phonon and electron-phonon coupling related quantities can be calculated. For example, with the Hellman-Feynman theorem it can be shown that from the linear response of the density $\Delta n$ the dynamical matrix can be constructed from which the phonon frequencies and modes can be computed. Further, from the first-order derivative of the self-consistent Kohn-Sham potential $\Delta V_{eff}$ the electron-phonon coupling matrix elements can be obtained

\begin{equation}
g_{{\bf k+q},{\bf k}}^{{\bf q}j,\nu \mu}=\bigg ( \frac{\hbar}{2\omega_{{\bf q}j}} \bigg )^{1/2} \langle \psi_{{\bf k+q}\mu} | \Delta V_{eff}^{{\bf q j}} | \psi_{{\bf k}\nu} \rangle,
\label{gmat}
\end{equation}
where $\omega_{{\bf q}j}$ refers to the phonon frequency corresponding to the phonon mode with wavevector ${\bf q}$ and irreducible representation $j$. From these electron-phonon coupling matrix elements and the phonon frequencies important quantities can be determined such as the spectral function (isotropic Eliashberg function) $\alpha^2F(\omega)$ and the isotropic coupling constant (electron-phonon mass-enhancement factor) $\lambda$, 
\begin{equation}
\begin{gathered}
\alpha^{2}F(\omega)=\frac{1}{N(\epsilon_F)}\sum_{\mu\nu}\sum_{{\bf q}j} \delta(\omega - \omega_{{\bf q}j}) \sum_{{\bf k}} |g_{{\bf k+q},{\bf k}}^{{\bf q}j,\nu \mu}|^2 \\
 \times \delta (\epsilon_{{\bf k+q}\mu} - \epsilon_F)  \delta(\epsilon_{{\bf k}\nu} - \epsilon_F), \\
\lambda = 2\int \frac{\alpha^{2}F(\omega)}{\omega}d\omega 
\end{gathered}
\label{quant}
\end{equation}
\newline
Here $N(\epsilon_F)$ is the density of states at the Fermi level indicated by $\epsilon_F$.

From temperature dependent scanning tunneling spectroscopy experiments on the Cr(001)-surface an electron-phonon mass-enhancement factor 5-10 times larger than in the bulk is predicted.~\cite{okon3} This prediction is obtained under the assumption that the resonance at the Fermi level is due to a $d_{z^2}$-surface state renormalized by electron-phonon coupling. Thus, this predicted $\lambda$ corresponds to the $d_{z^2}$ state of the surface layer. However, computationally a Cr(001)-surface is simulated by a finite number of layers. Then, the $\lambda$ calculated from Eq.~\ref{quant} corresponds in general to the whole system, i.e. not specifically to the top-layer. Therefore, a projection is required to obtain $\lambda$ corresponding to the top-layer and $d_{z^2}$ state. In order to obtain this projection it is instructive to first consider the $\psi_{{\bf k}\nu}$ dependence of the spectral function explicitly
\begin{equation}
\begin{gathered}
\alpha^{2}F_{{\bf k}\nu}(\omega)=\sum_{\mu}\sum_{{\bf q}j} \delta(\omega - \omega_{{\bf q}j}) |g_{{\bf k+q},{\bf k}}^{{\bf q}j,\nu \mu}|^2 \delta(\epsilon_{{\bf k+q}\mu} - \epsilon_F).
\end{gathered}
\label{afkv}
\end{equation}
\newline
Here $\alpha^{2}F(\omega)$ of Eq.~\ref{quant} is obtained by performing the following average over the Fermi surface \mbox{$\alpha^{2}F(\omega)=\frac{1}{N(\epsilon_F)}\sum_{{\bf k}\nu}\alpha^{2}F_{{\bf k}\nu}(\omega)\delta(\epsilon_{{\bf k}\nu} - \epsilon_F)$}. The projection of the $\alpha^{2}\widehat{F}(\omega)$ operator onto the local basis $|{\bf R}\xi \rangle$ (with ${\bf R}$ referring to the layer and $\xi=m\sigma$ containing both the orbital and spin projection) can be expressed in terms of Eq.~\ref{afkv} and the $\langle {\bf R}\xi| {\bf k}\nu\rangle$ coefficients,
\begin{equation}
\begin{gathered}
\langle {\bf R}\xi| \alpha^{2}\widehat{F}(\omega)| {\bf R'}\xi' \rangle =\sum_{{\bf k} \nu} \langle {\bf R}\xi| {\bf k}\nu\rangle \alpha^{2}F_{{\bf k}\nu}(\omega) \langle {\bf k}\nu | {\bf R'}\xi' \rangle,
\end{gathered}
\label{afproj}
\end{equation}
\newline
where $\alpha^{2}F_{{\bf k}\nu}(\omega)=\langle {\bf k}\nu| \alpha^{2}\widehat{F}(\omega)| {\bf k}\nu \rangle$. Then, by performing the appropriate summation and averaging over the Fermi surface the $|{\bf R}\xi \rangle$ dependence of $\alpha^{2}F(\omega)$ can be obtained 
\begin{equation}
\begin{gathered}
\alpha^{2}F_{{\bf R}\xi}(\omega)=\frac{1}{N(\epsilon_F)_{{\bf R}\xi}}\sum_{{\bf R'}\xi'} \langle {\bf R}\xi| \alpha^{2}\widehat{F}(\omega)| {\bf R'}\xi' \rangle \\
\times \langle {\bf R'}\xi' | \delta(\hat{H}_{DFT} - \epsilon_F) | {\bf R}\xi \rangle \\
= \frac{1}{N(\epsilon_F)_{{\bf R}\xi}} \sum_{{\bf k} \nu} \alpha^{2}F_{{\bf k}\nu}(\omega) \delta(\epsilon_{{\bf k}\nu} - \epsilon_F) |\langle {\bf R}\xi| {\bf k}\nu\rangle |^2.
\end{gathered}
\label{afrx}
\end{equation}
\newline
The expression in the last line of this equation is obtained by employing the unity operator $\sum_{{\bf k} \nu} |{\bf k} \nu\rangle \langle {\bf k} \nu |$ twice and inserting Eq.~\ref{afproj}. Here $N(\epsilon_F)_{{\bf R}\xi}$ is the projected density of states at the Fermi energy at the site with position vector ${\bf R}$ and of the state indicated by $\xi=m\sigma$. Further, $\alpha^{2}F(\omega)$ of Eq.~\ref{quant} is obtained by performing the summation $\frac{1}{N(E_F)}\sum_{{\bf R}\xi}\alpha^{2}F_{{\bf R}\xi}(\omega) N(\epsilon_F)_{{\bf R}\xi}$.

On its turn $\alpha^{2}F_{{\bf R}\xi}(\omega)$ of Eq.~\ref{afrx} can used to calculate the $|{\bf R}\xi \rangle$  projected electron-phonon mass-enhancement factor $\lambda_{{\bf R}\xi}$ and averaged $\lambda$ (of Eq.~\ref{quant}) via
\begin{equation}
\begin{gathered}
\lambda_{{\bf R}\xi} = 2\int \frac{\alpha^{2}F_{{\bf R}\xi}(\omega)}{\omega}d\omega,\\
\lambda=\frac{1}{N(\epsilon_F)}\sum_{{\bf R}\xi}\lambda_{{\bf R}\xi}N(\epsilon_F)_{{\bf R}\xi}.
\end{gathered}
\label{projlamb}
\end{equation}

The calculations in this work are performed by employing the DF(P)T implementation of the Quantum Espresso code.~\cite{espresso} An ultrasoft pseudopotential is used to reduce the required plane-wave kinetic energy cut-off with respect to norm-conserving pseudopotentials.~\cite{pseu1,pseu2} For details on the DFPT implementation with ultrasoft pseudopotentials we refer the reader to Ref.~\onlinecite{pseuphon}. Unless stated otherwise, the calculations are performed with an exchange-correlation functional in the generalized gradient approximation (GGA) as formulated by Perdew, Burke and Ernzerhof (PBE).~\cite{gga} For the Cr(001)-surface calculations a kinetic energy cut-off for the expansion into plane waves of the wavefunctions and density of respectively 70~Ry and 800~Ry were taken. In case of Cr-bulk calculations 50~Ry and 600~Ry were used. It was tested that the relevant quantities in this work, e.g. the electron-phonon mass-enhancement factors, are converged with respect to these energy cut-offs.   

From Eq.~\ref{quant} and~\ref{afrx} it can be seen that summations over electronic (${\bf k}$) and phononic (${\bf q}$) meshes are required. Since the electron-phonon coupling matrix elements are known to depend smoothly on ${\bf k}$ and ${\bf q}$, the interpolation scheme presented in Ref.~\onlinecite{malgor} is adapted. For the Cr(001)-surface calculations we tested that for a 15x15x1 to 30x30x1 interpolation of the ${\bf k}$-mesh and 5x5x1 to 10x10x1 interpolation of the ${\bf q}$-mesh convergence of the quantities of interest is achieved.  As for the Cr-bulk calculations an interpolation of the ${\bf k}$-mesh from 15x15x15 to 30x30x30 and for the ${\bf q}$-mesh from 5x5x5 to 10x10x10 was found to be adequate. 

Further, the calculations were scalar relativistic and spin-polarized, where an antiferromagnetic magnetic structure was taken. In addition for all DF(P)T calculations a geometry optimalization is performed such that the total energy and forces are converged to within $10^{-5}$~Ry and $10^{-4}$~a.u. For the Cr(001)-surface calculations a large vacuum of at least 20~\AA~was introduced to prevent interactions between layers of different unit cells. The Cr-bulk simulations were for the bcc structure. After geometry optimization a lattice constant of $2.87$~\AA~was obtained for an antiferromagnetic magnetic structure and $2.83$~\AA~for the paramagnetic situation.

\section{Results and Discussion}
\label{section-3}
Before the electron-phonon mass-enhancement factor of the Cr(001)-surface is investigated, first the Cr-bulk is addressed. In Table~\ref{tablbulk} the electron-phonon mass-enhancement factor (second column) and the density of states at the Fermi level $\epsilon_F$ per atom (third column) are presented for paramagnetic (PM) and antiferromagnetic (AFM) Cr-bulk. Further, both the LDA and the GGA exchange-correlation functional were employed. Since within LDA no stable AFM state could be obtained, the electron-phonon mass-enhancement factor could not be determined. The lack of a stable AFM solution within LDA is also observed in other studies.~\cite{okon6,afmun} However, for the paramagnetic state the difference between LDA and GGA appears to be small. Furthermore, from this table a large suppression in $\lambda$ of about a factor 3 can be observed for the antiferromagnetic structure compared to that of the paramagnetic case. Interestingly, this same factor also corresponds to the difference observed in the density of states at the Fermi level. Such a dependence of $\lambda$ on the density of states at the Fermi level can be inferred from Eq.~\ref{afkv} by assuming that the electron-phonon matrix elements ($g_{{\bf k+q},{\bf k}}^{{\bf q}j,\nu \mu}$) are approximately constant and that the phonon energies ($\omega_{{\bf q}j}$) can be neglected with respect to the electronic energies. Then, the term $\sum_{{\bf q}\mu}\delta(\epsilon_{{\bf k+q}\mu} - \epsilon_F)$ gives the proportionality with the density of states at the Fermi level. 

\begin{table}[h]
\begin{center}
\caption{The electron-phonon mass-enhancement factor $\lambda$ and density of states at the Fermi level per atom $\epsilon_F$ for paramagnetic (PM) and antiferromagnetic (AFM) Cr-bulk are presented. Here LDA and GGA refer to the employed exchange-correlation functional.}
\begin{tabular}{|l|c|c|}
\hline
 & $\lambda$ & $N(\epsilon_F)$  \\ \hline
PM (LDA)      & 0.37   &  4.7  \\ 
\hline
PM (GGA)      & 0.35   &  4.8  \\ 
\hline
AFM (GGA)   & 0.12   &  1.7 \\ 
\hline
\end{tabular}
\label{tablbulk}
\end{center}
\end{table}

The suppression of superconductivity in the antiferromagnetic phase for Cr-bulk is a known phenomenom.~\cite{optexp} Note that $\lambda$ is the central quantity in superconducting theory. Further, our calculated value of $\lambda$ for the antiferromagnetic phase is found to be in good agreement with optical pump-and-probe experiment, $\lambda=0.13\pm0.02$.~\cite{optexp} Unfortunately, the $\lambda$ obtained for the paramagnetic phase cannot be compared with experiment. However, strong coupling theory predicted a $\lambda$ of 0.25 for the paramagnetic phase, which is in reasonable agreement with our result.~\cite{strongt} It should be noted that the method we employed is much more sophisticated than strong coupling theory.  

\begin{figure}[!ht]
\begin{center}
\includegraphics[trim=80 30 30 60, clip, width=9cm, scale=0.5]{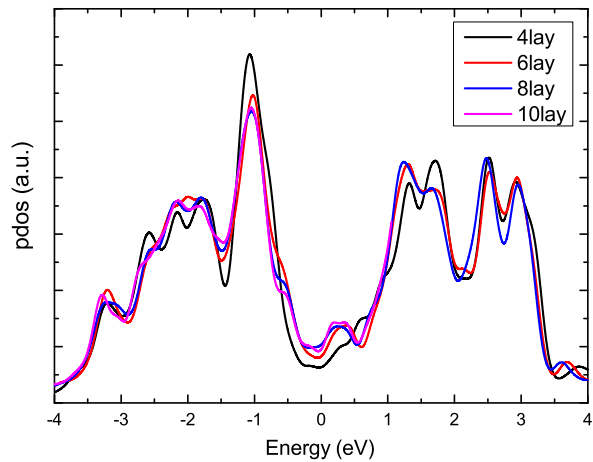}
\end{center}
\caption{The total 3d  projected density of states of the surface layer for calculations in which different number of layers are considered. Here black corresponds to a calculation of four layers, red to six layers, blue to eight layers and pink to ten layers.} 
\label{figpdos}
\end{figure}

In the following the electron-phonon mass-enhancement factor of the Cr(001)-surface is investigated. For this purpose it is instructive to first inspect the 3d projected density of states of the Cr(001)-surface, i.e. the top-layer of a system consisting of multiple layers. For this surface-layer the total 3d projected density of states is depicted in Fig.~\ref{figpdos} for calculations in which different number of layers are considered. Here black corresponds to a calculation of four layers, red to six layers, blue to eight layers and pink to ten layers. 
From Fig.~\ref{figpdos} it can be observed that for eight layers the projected density of states can be considered converged with respect to the total number of layers. Especially note that the important region for the electron-phonon mass-enhancement factor, which is around the Fermi level, is well converged for eight layers. Further, our computed 3d projected density of states is in good agreement with what is obtained in previous DFT studies.~\cite{okon6,schuler}

\begin{figure}[!ht]
\begin{center}
\includegraphics[trim=65 30 30 60, clip, width=9cm, scale=0.5]{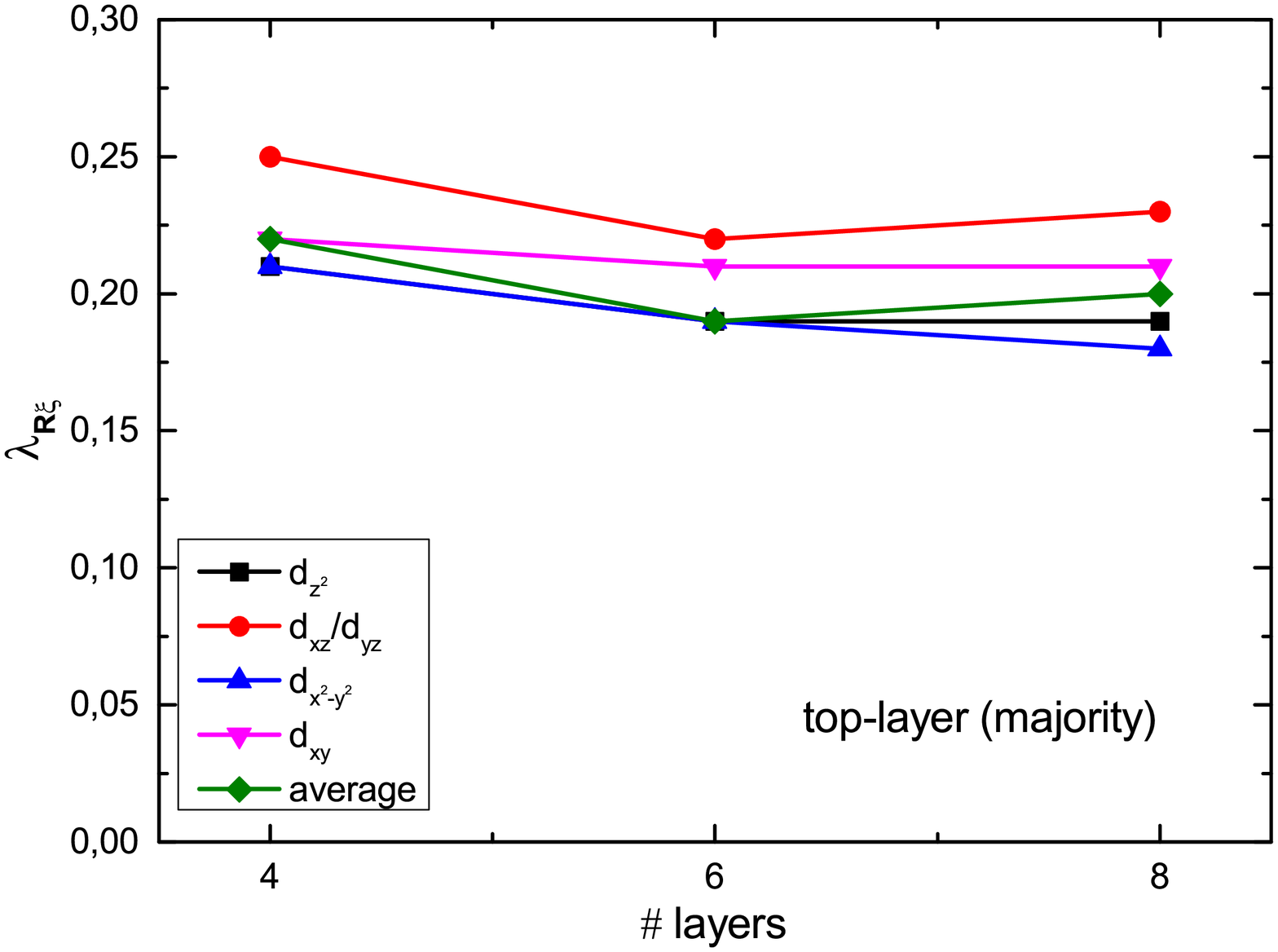}
\includegraphics[trim=65 37 30 60, clip, width=9cm, scale=0.5]{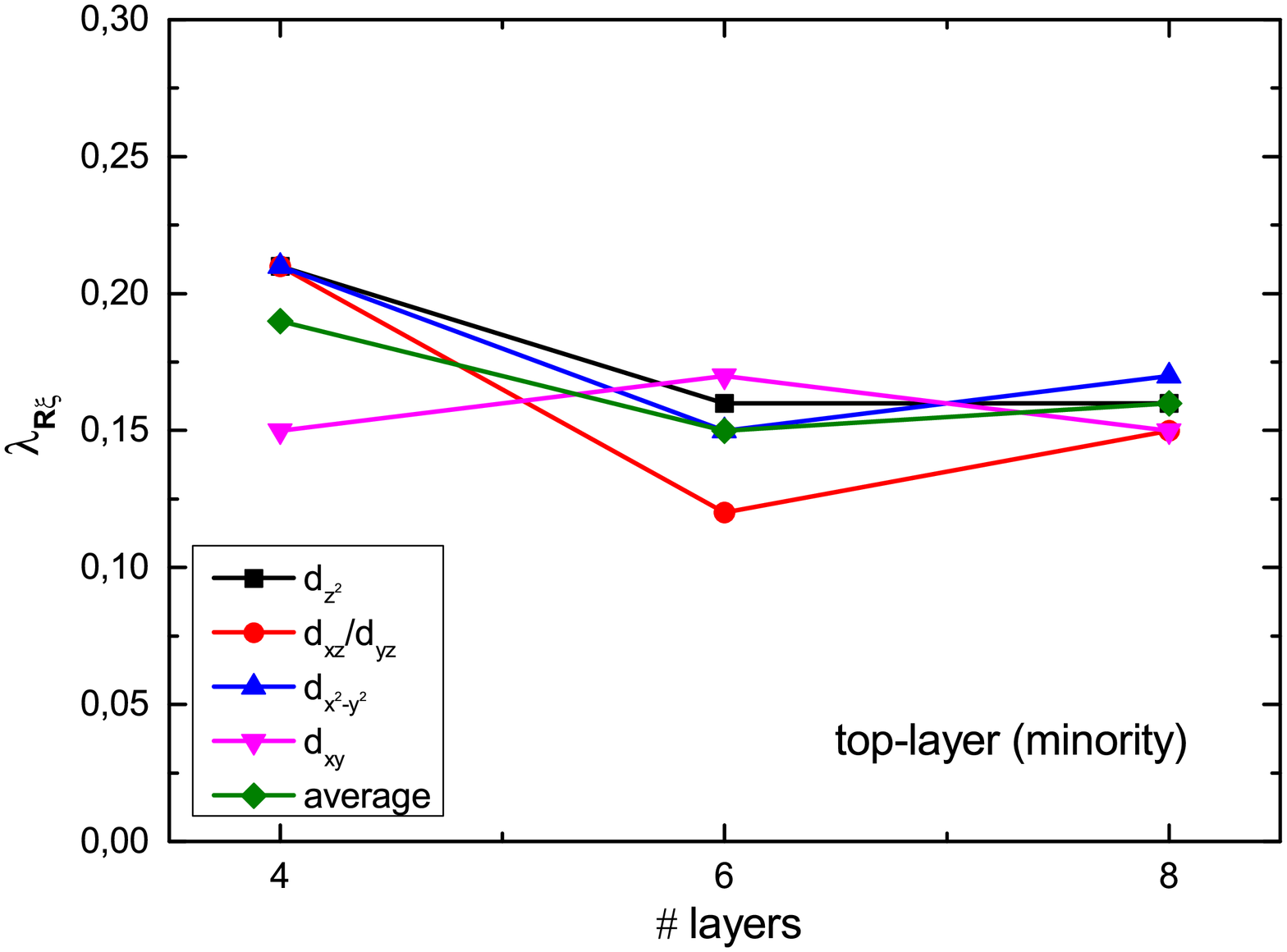}
\end{center}
\caption{The electron-phonon mass-enhancement factor of Eq.~\ref{projlamb} corresponding to the top-layer and the different 3d states is presented for calculations of different number of layers . Here the top figure referes to the majority spin states and the bottom figure to the minority spin states. Further, black corresponds to $d_{z^2}$, red to $d_{xz}/d_{yz}$, blue to $d_{x^2-y^2}$, pink to $d_{xy}$ and green to the spin averaged $\frac{1}{N(\epsilon_F)_{{\bf R}\sigma}}\sum_{m}\lambda_{{\bf R}m\sigma}N(\epsilon_F)_{{\bf R}\xi}$, where ${{\bf R}\xi}={\bf R}m\sigma$, $m$ is the sum over the different 3d-states and $N(E_F)_{{\bf R}\sigma}$ is the spin-projected and layer-projected density of states at the Fermi level.} 
\label{figprojlam1}
\end{figure}

In Fig.~\ref{figprojlam1} the electron-phonon mass-enhancement factor of Eq.~\ref{projlamb} corresponding to the top-layer and the different 3d states is presented for calculations of different number of layers. Here the top figure referes to the majority spin state and the bottom figure to the minority spin state. Further, black corresponds to $d_{z^2}$, red to $d_{xz}/d_{yz}$ (are equivalent due to symmetry at the surface), blue to $d_{x^2-y^2}$, pink to $d_{xy}$ and green to the per spin averaged $\frac{1}{N(E_F)_{{\bf R}\sigma}}\sum_{m}\lambda_{{\bf R}m\sigma}N(\epsilon_F)_{{\bf R}\xi}$, where ${{\bf R}\xi}={\bf R}m\sigma$, $m$ is the sum over the different 3d-states and $N(E_F)_{{\bf R}\sigma}$ is the spin-projected and layer-projected density of states at the Fermi level. From this figure it can be observed that for all majority spin 3d-states the electron-phonon mass-enhancement factor appears to be quite well converged with respect to the number of layers. For the minority spin 3d-states the fluctuation as function of the number of layers is a bit larger. This is especially the case for the $d_{xz}/d_{yz}$ states. However, the important observation is that for the (majority and minority) $d_{z^2}$ surface state the electron-phonon mass-enhancement factor seems to be converged already for six layers. Therefore, we consider $0.19$ and $0.16$ to be the electron-phonon mass-enhancement factor for respectively the majority and minority $d_{z^2}$ surface state of the Cr(001)-surface.

A further inspection of Fig.~\ref{figprojlam1} shows that on average the electron-phonon mass-enhancement factor for the majority 3d-states is a bit larger than for the minority 3d-states. The largest electron-phonon mass-enhancement factor is obtained for the $d_{xz}/d_{yz}$ majority states, while the smallest for the $d_{xz}/d_{yz}$ minority states. 

\begin{figure}[!ht]
\begin{center}
\includegraphics[trim=65 30 30 60, clip, width=9cm, scale=0.5]{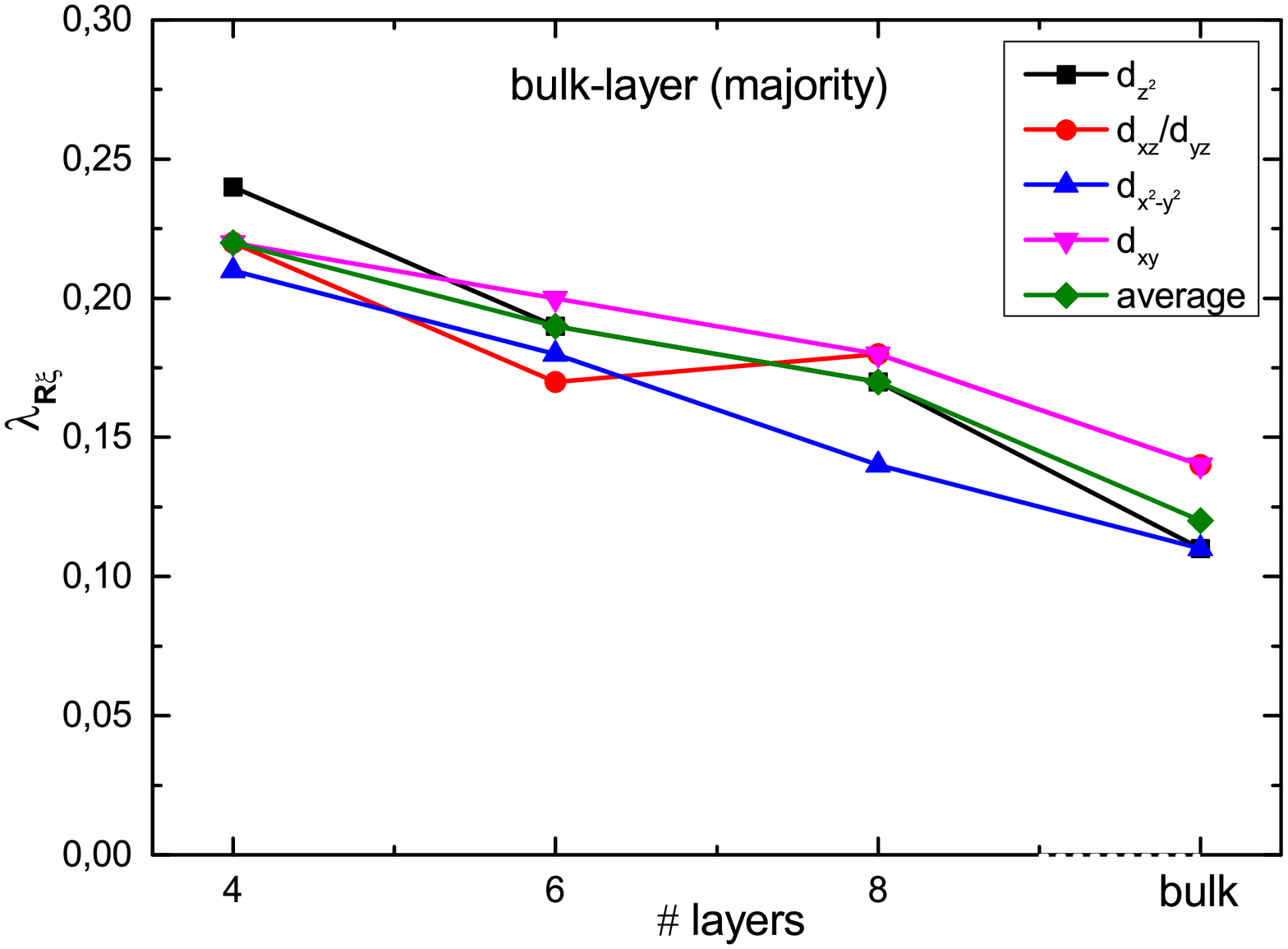}
\includegraphics[trim=65 37 30 60, clip, width=9cm, scale=0.5]{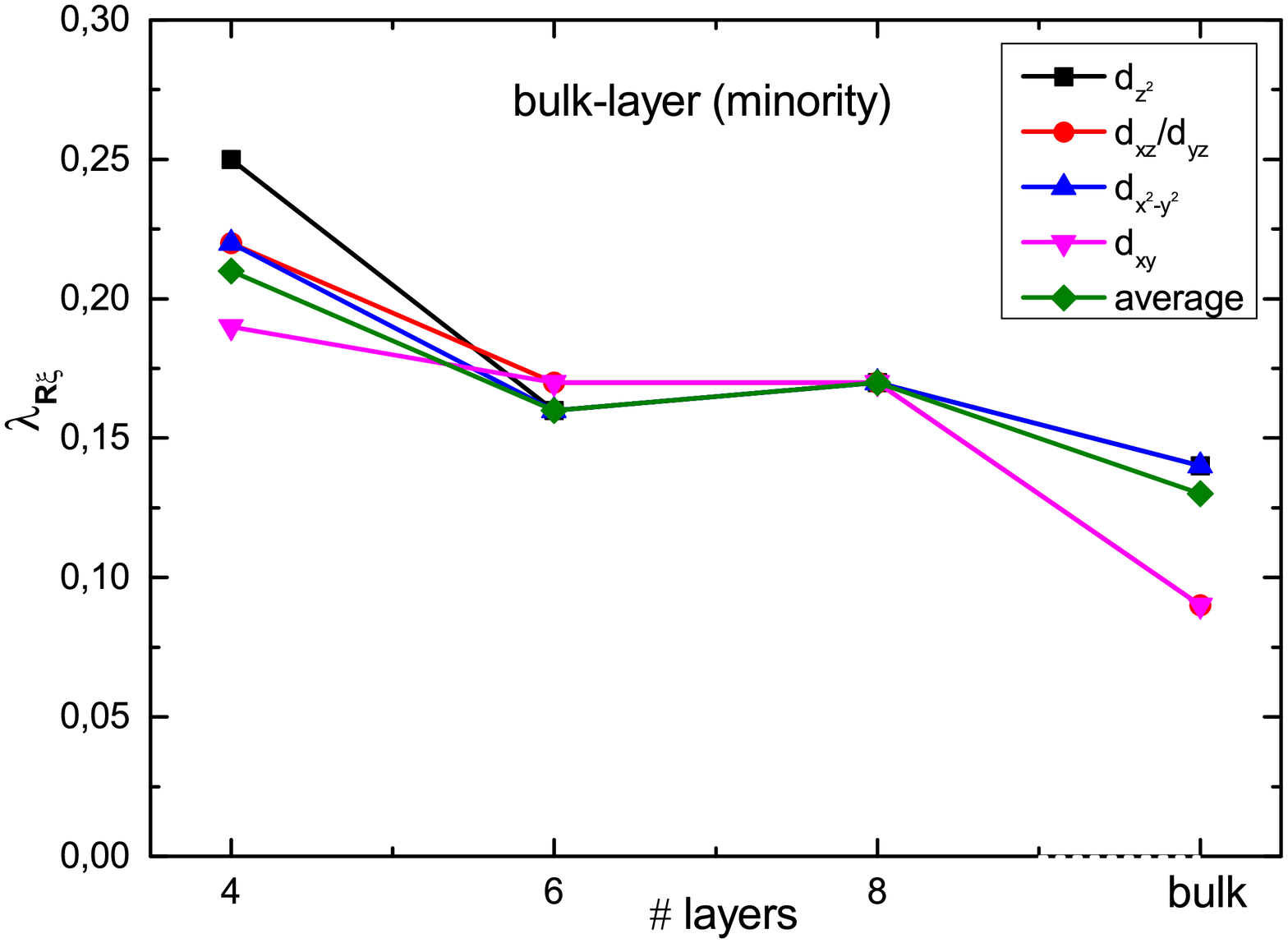}
\end{center}
\caption{The electron-phonon mass-enhancement factor of Eq.~\ref{projlamb} corresponding to the bulk-layer (middle layer) and the different 3d states is presented for calculations of different number of layers . Here the top figure referes to the majority spin states and the bottom figure to the minority spin states. Further, black corresponds to $d_{z^2}$, red to $d_{xz}/d_{yz}$, blue to $d_{x^2-y^2}$, pink to $d_{xy}$ and green to the spin averaged $\frac{1}{N(E_F)_{{\bf R}\sigma}}\sum_{m}\lambda_{{\bf R}m\sigma}N(\epsilon_F)_{{\bf R}\xi}$, where ${{\bf R}\xi}={\bf R}m\sigma$, $m$ is the sum over the different 3d-states and $N(E_F)_{{\bf R}\sigma}$ is the spin-projected and layer-projected density of states at the Fermi level.} 
\label{figprojlam2}
\end{figure}

\begin{figure}[!ht]
\begin{center}
\includegraphics[trim=80 37 30 60, clip, width=9cm, scale=0.5]{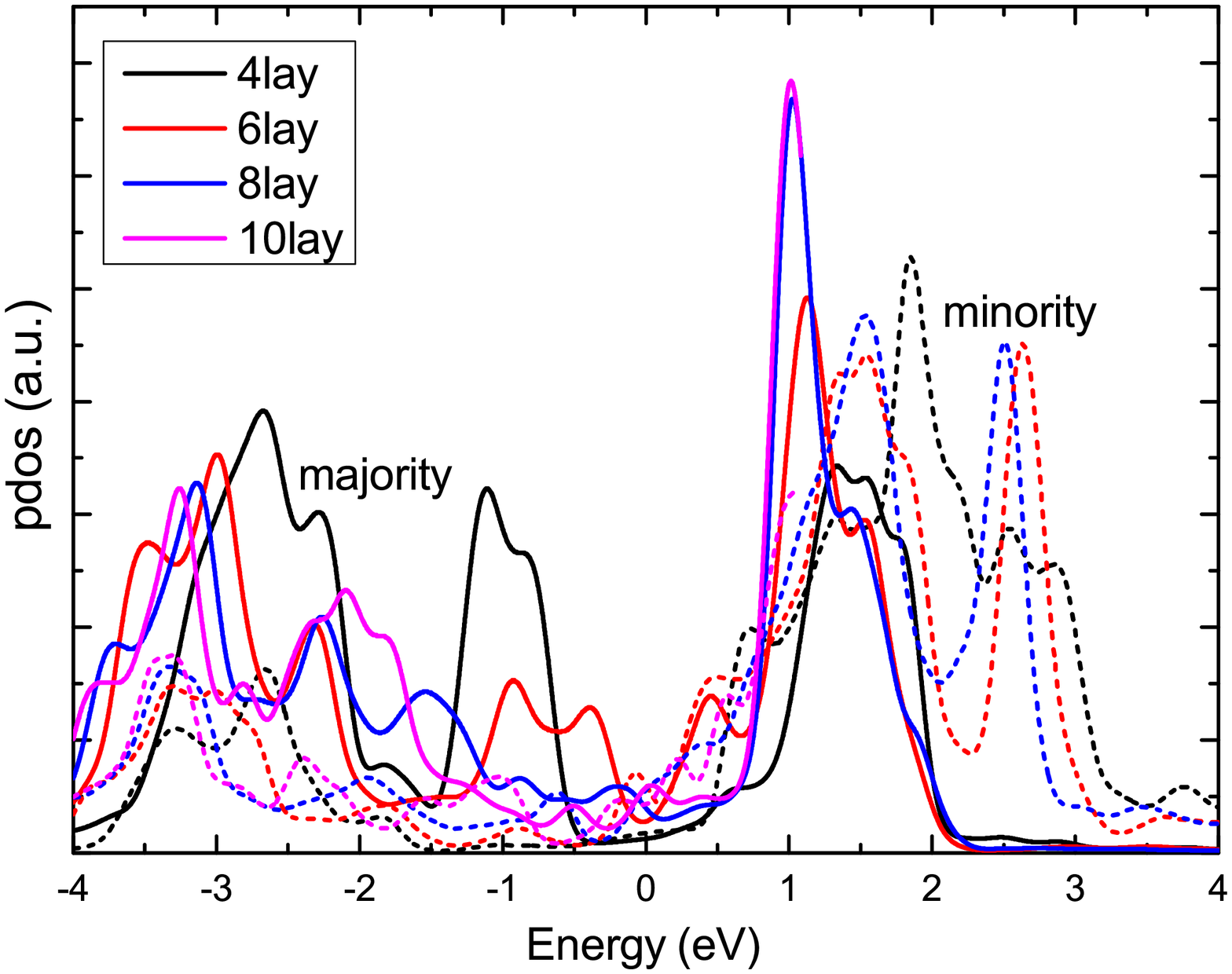}
\includegraphics[trim=80 37 30 60, clip, width=9cm, scale=0.5]{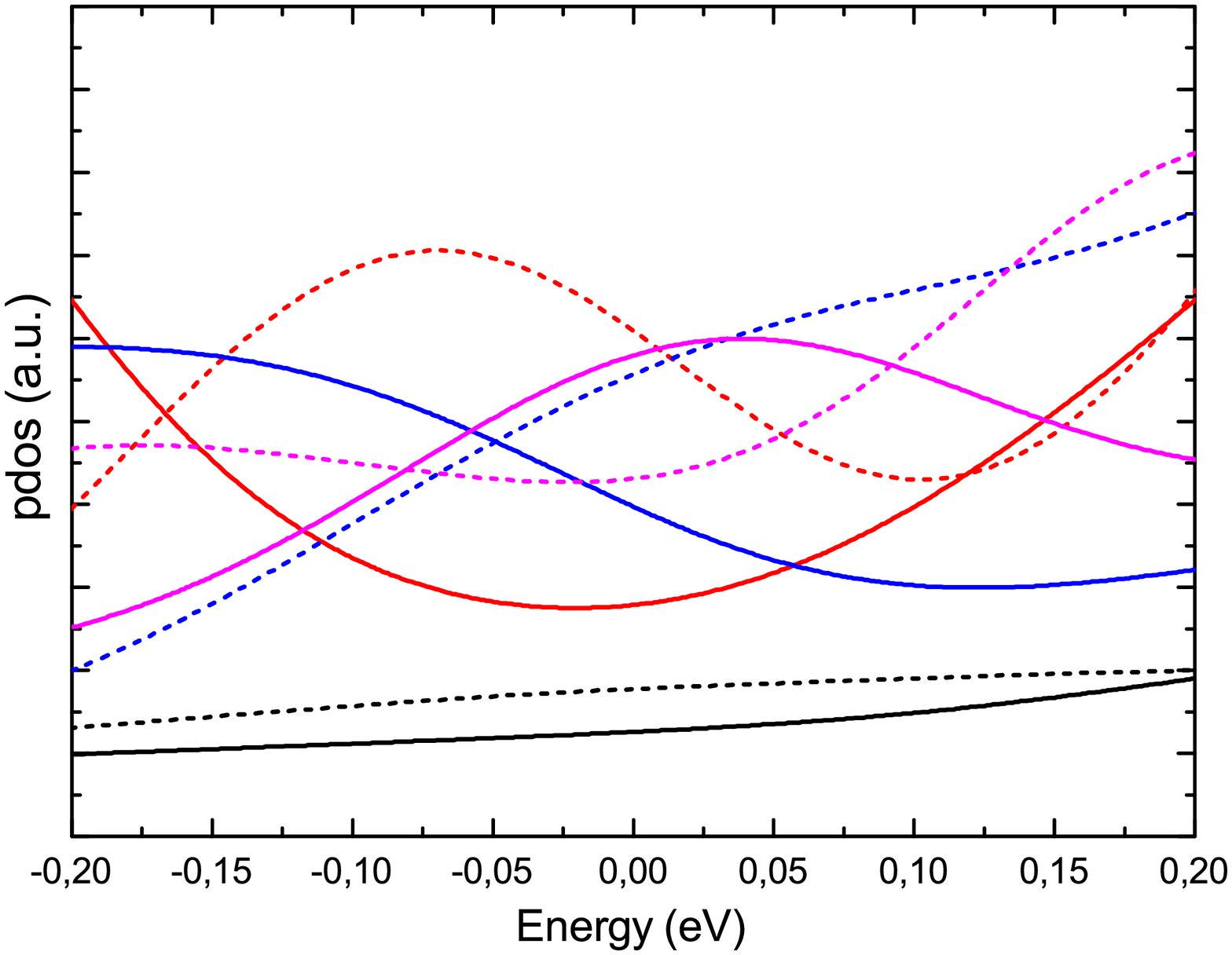}
\end{center}
\caption{The $d_{z^2}$ projected density of states of the bulk layer (middle layer) for calculations in which different number of layers are considered. Here black corresponds to a calculation of four layers, red to six layers, blue to eight layers and pink to ten layers. The solid lines refer to the majority spin channel and the dashed lines to the minority spin channel. The bottom figure is a zoom-in of the top figure around the Fermi level.} 
\label{figpdos2}
\end{figure}

It is also interesting to investigate the electron-phonon mass-enhancement factor of the 3d-states corresponding to the bulk-layer (middle layer). This is presented in Fig.~\ref{figprojlam2}, where also results for the Cr-bulk (indicated by 'bulk') are included. Note that for Cr-bulk due to symmetry the $d_{z^2}$ and $d_{x^2-y^2}$ states are equivalent, and also the $d_{xz}$, $d_{yz}$ and $d_{xy}$ states. Therefore, for these states the electron-phonon mass-enhancement factors should become equivalent as the number of layers increases. For eight layers the majority $d_{xz}$, $d_{yz}$ and $d_{xy}$ states are equivalent, even though the bulk value is not reached. This also occurs for the minority $d_{xz}$, $d_{yz}$ and $d_{xy}$ states, and $d_{z^2}$ and $d_{x^2-y^2}$ states. That the bulk-like values are not achieved for eight layers can be partly explained by the fact that the projected density of states at the Fermi level is not converged with respect to the number of layers. As a typical example, the $d_{z^2}$ projected density of states of the bulk-layer is presented as function of the number of layers in Fig.~\ref{figpdos2}. From this figure it is clear that for ten layers the projected density of states around the Fermi level is not converged with respect to the number of layers. 

As mentioned above the main interest is the electron-phonon mass-enhancement factor of the $d_{z^2}$-state corresponding to the top-layer. It was found that this quantity is converged already for six layers, which is in contrast to what was obtained for the bulk-layer. Therefore, the electron-phonon mass-enhancement factors of the majority and minority $d_{z^2}$ surface states, $0.19$ and $0.16$, are compared directly to those corresponding to the bulk of $0.11$ and $0.14$. There is less than a factor $2$ difference between them. The same holds when a comparison is made with the bulk spin averaged electron-phonon mass-enhancement factors of $0.12$ and $0.13$. On the other hand, experimentally a factor of about 5-10 was predicted. For convenience, the values of $0.19$ and $0.16$ can also be directly compared with the absolute value predicted in experiment, $1.53\pm0.40$.~\cite{okon3} It is clear that none of our calculated values are in agreement with this prediction. 

It should be noted that the missmatch with experiment is smaller, when realistic values for the temperature range and surface Debye frequency are employed for the fit of the experimental data. In order to understand this, the procedure to obtain the electron-phonon mass-enhancement factor in Ref.~\onlinecite{okon3} has to be inspected more closely. In their work the experimental data is fitted with the following model,
\begin{equation}
\begin{gathered}
\Gamma_{e-ph}(T)=\Gamma_{ee}+\lambda_{sur}\frac{2\pi}{\omega_D^2}\int_{0}^{\omega_D}dE'E'^2\big[1-f(E_{0}-E')\\
+2n(E')+f(E_{0}+E')\big].
\end{gathered}
\label{model}
\end{equation}
Here $\omega_D$ is the Debye frequency corresponding to the surface, $T$ the temperature, $\lambda_{sur}$ is the electron-phonon mass-enhancemant factor of the $d_{z^2}$-surface state, $E_{0}$ is the position of the resonance, $f(E)$ the Fermi distribution and $n(E)$ the Bose-Einstein distribution. Further, $\Gamma_{e-ph}$ is the inverse life-time due to electron-phonon processes within the Debye model, and $\Gamma_{ee}$ is the inverse life-time due the electron-electron interactions. The latter can be approximated by a constant for energies close to the Fermi level and low enough temperatures, i.e. it then corresponds to the off-set observed at zero temperature.

The first thing we noticed is that in Ref.~\onlinecite{okon3} $\lambda_{sur}$ is assumed to be constant, while the model of Eq.~\ref{model} is used to fit the data in a temperature range of $4-350$~K. The problem is that in this temperature range the paramagnetic to antiferromagnetic phase transition is crossed at $311$~K. From for example Table~\ref{tablbulk} it is clear that the electron-phonon mass-enhancement factor can change drastically between these regimes. Therefore, we performed a fitting in the temperature range $4-178$~K with Eq.~\ref{model} to obtain $\Gamma_{ee}$, $\lambda_{sur}$ and $\omega_D$, where we took $E_{0}=20\pm5$~meV (same as in Ref.~\onlinecite{okon3}). It appears then that the error bars of the experimental data are too large to accurately predict these quantities, $\Gamma_{ee}=14\pm9$, $\lambda_{sur}=1.50\pm1$ and $\omega_D=48\pm25$. This is the reason why in Ref.~\onlinecite{okon3} (and also in Ref.~\onlinecite{rehbein}) only two quantities are fitted at a time and for the other a 'reasonable' guess is taken. More precisely, in Ref.~\onlinecite{okon3} first the Debye frequency $\omega_D$ corresponding to the surface is taken to be equal to the bulk, $52.5$~meV, for the determination of $\Gamma_{ee}$ and $\lambda$. However, for the Cr(001)-surface the Debye frequency has been found both experimentally and theoretically to be at least 2 times smaller than in the bulk.~\cite{wdexp,jack} On the other hand, $\Gamma_{ee}=19.5\pm5$~meV seems to be a reasonable approximation for the off-set observed at zero temperature in Fig.~3(b) of Ref.~\onlinecite{okon3}. Then, taking $\omega_D=26$~meV (half of the bulk value) and $\Gamma_{ee}=19.5\pm5$~meV (and $E_{0}=20\pm5$~meV) we find for $\lambda_{sur}$ by employing Eq.~\ref{model} in the temperature range $4-178$~K the following, $\lambda_{sur}=0.77\pm0.16$. It is clear that this value is not in agreement with the calculated electron-phonon mass-enhancement factors of the majority and minority spin surface $d_{z^2}$-states ($0.19$ and $0.16$).  

We could also use $\Gamma_{ee}=19.5\pm5$~meV and $E_{0}=20\pm5$~meV to find both $\lambda_{sur}$ and $\omega_D$ (corresponding to the surface) by employing Eq.~\ref{model} in the temperature range $4-178$~K. Then, $\lambda=1.44\pm0.83$ and $\omega_D=48\pm25$ are found. From these results it appears again that the error bars of the experimental data points are too large to accurately determine these quantities. Therefore, we are convinced that it is better to take $\omega_D=16$~meV from experiment~\cite{wdexp}. Then, employing Eq.~\ref{model} with $\Gamma_{ee}=19.5\pm5$~meV and $E_{0}=20\pm5$~meV in a temperature range of $4-178$~K results in $\lambda_{sur}=0.70\pm0.10$, which is not in agreement with our calculations.

Besides the difference of about a factor $3$ between our calculations and the experimental prediction, there is a subtlety that should be addressed in more detail. This subtlety is related with the interpretation of the experimentally observed resonance at the Fermi level of Cr(001)-surfaces in terms of a $d_{z^2}$-surface state renormalized by electron-phonon coupling. Namely in absence of the electron-phonon coupling, i.e. at zero temperature, there should still be a peak at the Fermi level with a width determined by electron-electron interactions ($\Gamma_{ee}$) and impurities. However, from our DFT calculations it is clear that there is no such $d_{z^2}$-peak in the vicinity of the Fermi level (see the bottom figure of Fig.~\ref{figpdos}). The closest peak is at about $0.2$~eV with a width of roughly $0.25$~eV. This position is too far from the Fermi level and the width too large for the peak to allow for a substantial renormalization by electron-phonon coupling. For a substantial renormalization by means of the electron-phonon coupling it is known that both the position (with respect to the Fermi level) and the width of the peak can be at most of the order of the Debye frequency.~\cite{hewson} In an earlier work it was proposed that the overestimation of the magnetic moment within DFT was responsible for the wrong peak position.~\cite{okon6} However, it was immediately found that a correct peak position would result in an unrealistic underestimation of the magnetic moment.~\cite{okon6} In addition, the too large width of the peak could not be resolved by this. Another reason for the lack of a $d_{z^2}$ peak with a correct width at the Fermi level could be an insufficient treatment of electron-electron correlation effects. In recent DFT+DMFT calculations within the CTQMC solver, it has been demonstrated that depending on the double counting a $d_{z^2}$-peak can emerge at the Fermi level (see Fig.~3(d) of Ref.~\onlinecite{schuler}). 

Based on our findings presented in this work we conclude that the experimentally observed resonance at the Cr(001)-surface is not due to the renormalization of the $d_{z^2}$-surface state by means of electron-phonon coupling. Instead, electron-electron correlation effects are responsible for the experimentally observed resonance at the Fermi level. More studies are needed to precisely determine which correlation effects are important.


\section{Conclusion}
\label{section-4}
We have performed density functional perturbation theory calculations within a pseudopotential plane-wave approach to investigate the electron-phonon mass-enhancement factor of Cr(001)-surfaces and Cr-bulk. For Cr-bulk we made the interesting observation that within the paramagnetic phase the electron-phonon mass-enhancement factor is about $3$ times larger than in the antiferromagnetic phase. The same difference is found in the density of states at the Fermi level, which explains this behavior. Further, for the antiferromagnetic phase the calculated electron-phonon mass-enhancement factor is found to be in good agreement with experiment, while for the paramagnetic phase a reasonable agreement with strong coupling theory is obtained. For the paramagnetic phase it was also found that there is only a small difference between the electron-phonon mass-enhancement factor obtained in LDA and GGA.  

For the Cr(001)-surface we obtained for the majority and minority spin \textit{$d_{z^{2}}$} surface states an electron-phonon mass-enhancement factor of respectively $0.19$ and $0.16$. Compared to the bulk these are less than a factor $2$ larger, while experimentally a factor $5-10$ difference was predicted. Further, we showed that the difference between experiment and our calculations is smaller if we use realistic values for the temperature range and surface Debye frequency to fit the experimental data. More precisely, then experimentally an electron-phonon mass-enhancement factor of $0.70\pm0.10$ is obtained, which is not in agreement with our calculated values of $0.19$ and $0.16$. 

Based on these findings we conclude that the experimentally observed sharp resonance at the Fermi level of the Cr(001)-surface is not due to polaronic effects, i.e. is not due to a $d_{z^2}$ surface state renormalized by electron-phonon coupling. Instead, electron-electron correlations effects are responsible for the occurence of the resonance. More studies are needed to specify exactly which many-body processes this are.    

 
\subsection*{Acknowledgements}
The Nederlandse Organisatie voor Wetenschappelijk Onderzoek (NWO) and SURFsara 
are acknowledged for the usage of the LISA supercomputer and their support. L.P. and
M.I.K. acknowledge a support by European ResearchCouncil (ERC) Grant No. 338957. A.N.R. acknowledges support from the Russian Science Foundation, Grant 17-72-20041.



\end{document}